\documentstyle[aps,12pt,epsfig]{revtex}

\newcommand{\bm}{\bibitem}

\newcommand{\cp}{\chi^{(+)}}

\newcommand{\vv}{V_{bc}({\bf r}_1)}

\newcommand{\ro}{{\bf r}_1}
\newcommand{\ak}{{\bf k}_a}
\newcommand{\bq}{{\bf k}_b}

\newcommand{\cq}{{\bf k}_c}

\tighten

\begin{document}

\draft

\title{Coulomb and nuclear breakup effects in the single neutron removal
reaction $^{197}$Au($^{17}$C,$^{16}$C$\gamma$)X}
\date{\today}
\author{V. Maddalena$^a$, and R. Shyam$^{a,b}$}
\address { 
$^a$ National Superconducting Cyclotron Laboratory, Michigan State University,
East Lansing, MI 48824, U.S.A. \\ 
$^b$ Saha Institute of Nuclear Physics, Calcutta 700064, India} 
\maketitle

\begin{abstract}
We analyze the recently obtained new data on the partial cross sections
and parallel momentum distributions for transitions to 
ground as well as excited states of the $^{16}$C core, in the 
one-neutron removal reaction $^{197}$Au($^{17}$C,$^{16}$C$\gamma$)X
at the beam energy of 61 MeV/nucleon. The Coulomb and nuclear breakup
components of the one-neutron removal cross sections have been
calculated within a finite range distorted
wave Born approximation theory and an eikonal model, respectively.
The nuclear contributions dominate the partial cross sections
for the core excited states. By adding the nuclear and Coulomb 
cross sections together, a reasonable agreement is obtained with the 
data for these states. The shapes of the experimental 
parallel momentum distributions of the core states are 
described well by the theory. 
\end{abstract}
\newpage

The usefulness of the single nucleon transfer reactions in probing the
single-particle structure of the stable nuclei, is well 
established (see e.g. \cite{sat83,aus70,fes92,gle83}).
The theory of these reactions developed within the framework of
the distorted wave Born approximation (DWBA), has been
widely used to analyze the absolute magnitudes and shapes of the
measured cross sections to make the angular momentum assignments,
and deduce the spectroscopic factors for the ground as well as excited
states of the residual nuclei. However, the transfer reactions are not
yet routinely used  in probing the structure of exotic nuclei near the
neutron and proton drip lines. Although the first theoretical \cite{len98}
and experimental \cite{win00} studies of the feasibility of such
investigations have already been reported, some
formidable difficulties (see e.g. \cite{len98}) still persist in
the application of this method to probe the exotic nuclei. 
  
Recently, an alternative new and more versatile technique to investigate
the spectroscopy of nuclei near the drip line has been developed
\cite{nav98,aum00,nav00,gui00,val00}. In this method one nucleon 
(usually the valence or halo one) is removed from the projectile (a)
in its breakup reaction in the field of a target nucleus.
The states of the core (b) populated in this reaction are
identified by their gamma ($\gamma$) decays, whose intensities are
used to measure the partial breakup cross sections for these states.
The signatures of the orbital angular momentum
($\ell$) associated with the relative motion of different core states with
respect to the valence nucleon (removed from the projectile),
are provided by the corresponding parallel momentum distributions
also measured in this experiment. 
 
The practical experimental advantages of this method, such as  
large partial cross sections for the excitation of  
various bound states of the core fragment, the possibility of using 
thick targets, and the strong forward focusing of the reaction products,
make it possible to work with the high energy projectiles of low beam
intensities. This is in contrast with the existing 
situation in the case of transfer reactions. Furthermore, while in the
latter case the angular distributions of the ejectile loose their
characteristic $\ell$ dependence at high energies \cite{bou81}, the
parallel momentum distributions of the core states in the breakup
reactions still show strong dependence on the $\ell$ value
\cite{han96}. So far, most of the the studies of the (a,b$\gamma$)
type of reaction have been reported for the $^{11}$Be \cite{aum00},
$^{12}$Be \cite{nav00}, $^{14}$B \cite{gui00}, and $^{16,17,19}$C
\cite{val00} projectiles on a light $^9$Be target. Therefore, for
these cases the breakup process
is governed almost entirely by only the nuclear interaction between
the projectile fragments and the target; the Coulomb breakup contributions 
are almost negligible for these reactions. 

Using the framework of the post form distorted wave Born approximation,
a theory for the Coulomb breakup reactions has recently  
been developed \cite{cha00}. The finite range effects are included
in this theory, which can be applied to projectiles of any core
fragment-valence neutron angular momentum structure. This theory has
been applied rather successfully to investigate the inclusive data
for the breakup of halo nuclei on heavy targets at beam energies below
100 MeV/nucleon \cite{cha00}. A recent study within this theory 
\cite{shp00} of the A(a,b$\gamma$)X type of reaction involving
halo projectile nuclei on a $^{208}$Pb target, reveals that
the characteristics of this reaction are different
in a Coulomb dominated process as compared to those in the
nuclear dominated one. In the former case, transitions to the
excited states of the core fragment are found to be very weak,
which has been confirmed in a recent measurement of the
$^{197}$Au($^{14}$B,$^{13}$B$\gamma$)X reaction
\cite{gui00}. The pure Coulomb breakup cross sections decrease strongly
with the increasing separation energy and the $\ell$-value
of the core-valence neutron relative motion. 

In this paper, our aim is to investigate the one-neutron removal 
reaction of the (a,b$\gamma$) type induced by a non-halo nucleus
on a heavy target. We would like to see if the predictions of the
pure Coulomb breakup reaction in this case are different from those
described above. In this context, the $^{17}$C is interesting
in many respects. The ground state of this nucleus has a spin parity
of ${\frac{3}{2}^+}$ \cite{val00,war92}, which means that  
the relative motion of the $^{16}$C(g.s.)-valence neutron system
has a $\ell$ value of 2. This makes it an unlikely candidate for
having a halo structure even though the corresponding one-neutron
separation energy (SE) is only 0.729 MeV. Due to its non-halo
nature, the breakup of this nucleus is expected to occur in regions
around the distance of closest approach. Therefore, nuclear breakup cross
sections are likely to be important for this case even if the measurements
are performed on a heavy target. Moreover, the excited bound states
of $^{16}$C can have configurations in which the relative motion
between the excited core fragment and the valence neutron has a $\ell$ value
of zero. This implies that the partial cross sections for  
these states of $^{16}$C may have larger values even in a Coulomb
dominated breakup process.  

Motivated by these facts, we undertook the analysis the
$^{197}$Au($^{17}$C,$^{16}$C$\gamma$)X reaction at the beam 
energy of 61 MeV/nucleon, which was measured
in the same experiment in which the data were taken on a $^9$Be target
\cite{val00}. The detailed description of the technique and data 
analysis is presented in \cite{val00}. The Doppler
corrected $\gamma$-ray spectrum from the decay of the $^{16}$C residues
produced in this reaction, is shown in Fig. 1.
It shows feeding to the same states as those seen in the experiment with
a $^9$Be target. The partial cross sections (PCS) for transitions to these 
states, extracted from the absolute gamma branching ratios, are shown
in table I. It can be seen that the PCS to the 
excited states are quite substantial, which is contrary to the 
results of the measurements performed \cite{gui00} with $^{14}$B (a 
one-neutron halo nucleus) on the same target and beam energy where no
transition to the excited core states (of $^{13}$B) was observed.
 
In the theoretical analysis of the data, we assume
that the nuclear and Coulomb breakup cross sections (calculated with
different theories) can be added and that the Coulomb-nuclear
interference term can be neglected. Since this reaction is essentially
inclusive in nature (as the measurements are performed only for the heavy
fragment), the nuclear partial cross sections (NPC) have
contribution from both elastic (also known as diffraction dissociation)
and inelastic (also known as stripping or breakup-fusion) breakup modes
\cite{baur84,kasa82}. Cross sections for both these modes  
were calculated \cite{tos99} within an eikonal model
\cite{yab92,ber96,bar96,evl86} where the core-target and neutron-target
interactions are treated in the black disc approximation
and the optical limit of the Glauber theory \cite{tos99,alk96,glau57},
respectively. The data of Refs. \cite{nav98,aum00,nav00,gui00,val00} have
been analyzed within this model. For the semiclassical methods to calculate
the nuclear breakup cross sections, we refer to \cite{angela}.
 
The pure Coulomb breakup cross sections have been calculated by using a 
theory formulated \cite{cha00} within the framework of the post form
distorted wave Born approximation (DWBA). Within this theory,
the triple differential cross section for the reaction,
$ a + t \rightarrow b + c + t $, where $a$ is the projectile, $t$ the target,
and  $b$ (charged core) and $c$ (valence neutron) are the breakup fragments
in the final channel, is given by  
\begin{eqnarray}
{{d^3\sigma}\over{dE_bd\Omega_bd\Omega_c}} & = & 
{2\pi\over{\hbar v_a}}\rho(E_b,\Omega_b,\Omega_c)
\sum_{\ell m}|\beta_{\ell m}|^2,
\end{eqnarray}
where $\rho(E_b,\Omega_b,\Omega_c)$ is the appropriate 
\cite{cha00} three-body phase space factor. The reduced amplitude
$\beta_{\ell m}$ is defined as, 
\begin{eqnarray}
&&{\hat \ell}\beta_{\ell m}= 
 Z_\ell \int d {\bf r} \chi^{(-)*}_b(\bq,{\bf r})
e^{-i\delta\cq \cdot {\bf r}} \cp_a(\ak,{\bf r}),
\end{eqnarray}
where
\begin{eqnarray}
Z_\ell & = & \int d{\bf r}_1 e^{-i(\gamma\cq - \alpha {\bf K}) \cdot \ro}
V_{bc}(\ro) u_\ell(r_1) Y_{\ell m}({\hat r}_1),\\
\hat{\ell} & = & \sqrt{2\ell + 1}.
\end{eqnarray} 
In Eq. (2) $\chi's$ are the distorted waves for relative motions of
the center of mass (c.m.) of $a$ and $t$ and fragment $b$ and $t$, 
respectively. $\ak$, $\bq$, and $\cq$ are the Jacobi wave vectors
associated with the relative motions of $a$, $b$ and $c$, respectively.
The charged fragment $b$ interacts with the target by a point Coulomb
interaction, and hence $\chi^{(-)}_b({\bq},{\bf r})$ is a Coulomb
distorted wave with incoming wave boundary condition. The structure
function $Z_\ell$ involves the radial part of the wave function ($u_\ell$)
for the relative motion of the $b$-$c$ system and the corresponding
interaction $\vv$. For further theoretical details and definitions of 
other variables, we refer to \cite{cha00}. It may be noted that 
Eq. (2) treats the interaction $V_{bc}$ to all orders.
\begin{figure}
\begin{center}
\mbox{\epsfig{file=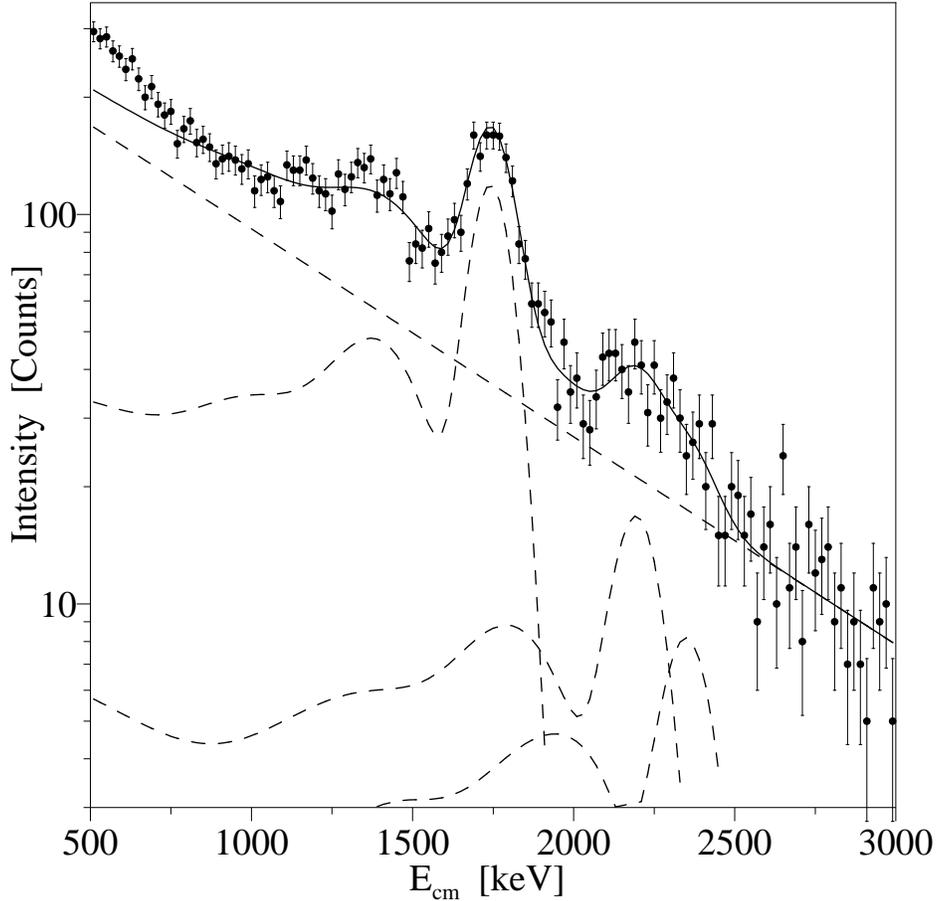,height=12.0cm}}
\end{center}
\vskip .3in
\caption{
Doppler-corrected $\gamma$-ray spectrum measured in
$^{197}$Au($^{17}$C, $^{16}$C+$\gamma$)X.  The black curve is a fit to
the spectrum using a single exponential curve for the background
and Monte-Carlo simulated response
functions (dashed curves) for each of the $\gamma$-ray transitions.
The spectrum was fitted using the procedure described in
\protect\cite{val00}.} 
\label{fig:figa}
\end{figure}
\noindent

An alternative theory of Coulomb breakup has also been developed
\cite{tos98} within the framework of an adiabatic model.
The expressions for the breakup amplitudes, within 
this theory, are similar to those of the finite range DWBA theory, 
although the two have been obtained from quite different assumptions.
In the studies of breakup reactions carried out so far, the two 
theories produced almost identical results in most of the cases \cite{cha00}.  
  
In calculations of both nuclear and Coulomb partial cross sections 
for populating a given final core fragment state, the projectile
ground state is described as having a configuration in which a
valence nucleon, with single particle quantum numbers
($n\ell j$) (see e.g. ref \cite{cha00}) and an associated
spectroscopic factor ($C^2S$), is coupled
to the specific core state ($I^\pi$). The total cross section in
each case is the sum \cite{nav98,tos99} of the cross sections
calculated with configurations
(having non-vanishing spectroscopic factors) 
corresponding to all the allowed values of the
channel spin. 

We assume the ground state (0$^+$) of $^{16}$C to arise from  
the removal of the valence neutron from the configuration 
($0d_{3/2} \otimes 0^+$, SE = 0.729 MeV) of the $^{17}$C ground state.
For the excited $2^+$ state at 1.77 MeV, two configurations,  
($0d_{5/2} \otimes 2^+$, SE = 2.499 MeV) and
($1s_{1/2} \otimes 2^+$, SE = 2.499 MeV) are considered. For the group 
of excited states near 4.1 MeV (2,3$^{(+)}$,4$^+$), we assume the
configurations, ($0d_{5/2} \otimes I^\pi$, SE = 4.829 MeV) and
($1s_{1/2} \otimes I^\pi$, SE = 4.829 MeV). The
corresponding spectroscopic factors ($C^2S$) are taken from \cite{war92}.
In each case, the neutron single particle wave function has been
calculated in a central Woods-Saxon well of radius 1.25 fm and diffuseness
0.7 fm. The depth of this well is adjusted to reproduce the corresponding
value of SE. 

Our results for the partial cross sections are shown in table I. 
It is evident that the theoretical partial cross sections 
(even the pure Coulomb breakup ones) to the excited states are
quite substantial.  This is in sharp contrast to
the results seen in the case of such reaction studied with 
halo nuclei having a $s$-wave core(g.s.)-neutron relative
motion in their ground states. It is interesting
to note that the nuclear partial cross sections are quite large.
For the ground state of $^{16}$C, NPC is of the similar magnitude as the
corresponding Coulomb
\begin{table}[here]
\begin{center}
\caption {Calculated partial cross sections to the final
states of $^{16}$C in the one-neutron removal reaction of 
$^{17}$C on a $^{197}$Au target at the beam energy of 61 MeV/nucleon.
$\sigma_C$ and $\sigma_N$ represent the partial cross sections due to
Coulomb and nuclear breakup processes. The latter is the sum of the
cross sections obtained in diffraction dissociation ($DD$) and stripping
($str$) mechanisms. $I^\pi$ and $E_x$ represent the spin-parity and
excitation energy of the core states respectively.}
\vspace{1.1cm}
\begin{tabular}{|c|c|c|c|c|ccc|c|c|c|}
I$^\pi$ &$E_x$ & $\ell$ & $C^2S$ & $\sigma_C$ & $\sigma_N^{DD}$&
$\sigma_N^{str}$&$\sigma_N$ & $\sigma_{th}$
& $C^2S \cdot \sigma_{th}$ & $\sigma_{exp}$ \\
        &      &        &        &            &                &
        &      & ($\sigma_C+\sigma_N$) &        &                \\
         &(\footnotesize{MeV})  &        &  & (\footnotesize{mb}) &
(\footnotesize{mb}) &(\footnotesize{mb}) & (\footnotesize{mb}) &
(\footnotesize{mb})& (\footnotesize{mb}) &
(\footnotesize{mb})\\
\hline
0$^+$       & 0.0   & 2 & 0.03 &148 & 38 & 75 &113 &261 &  8 & 113$\pm$26\\
\hline
2$^+$       & 1.77  & 0 & 0.16 &110 & 57 &106 &163 &273 & 44 &          \\
            &       & 2 & 1.44 & 32 & 22 & 52 & 74 &106 &153 &          \\
            &       &sum&      &142 & 79 &158 &237 &379 &197 & 162$\pm$56\\ 
\hline
2,3$^{(+)}$4$^+$& 4.1   & 0 & 0.22 & 32 & 31 & 69&100 &132 & 29  &           \\
            &       & 2 & 0.76 & 10 & 15 & 40& 55 & 65 & 49  &           \\
            &       & sum&     & 42 & 46 &109&155 &197 & 78  &  75$\pm$25 \\
\end{tabular}
\end{center}
\end{table}
\noindent
partial cross section (CPC). However,
for the excited states, the NPC dominates the CPC.
This can be understood from the fact that Coulomb breakup cross
sections decrease strongly as the value of SE increases,
while nuclear breakup cross sections have a weaker SE dependence. 
Furthermore, as expected the cross sections
to $\ell$ = 0 states are larger in comparison with those with 
$\ell$ = 2.

The sum of the NPC and CPC is in reasonable agreement with the
data for the excited states. However, for the ground state, the theoretical
partial cross section is more than an order of magnitude smaller than
the corresponding experimental value. Similar observation was also
made in the analysis of the experimental data on a $^9$Be target  
\cite {val00}. A $J^\pi$ assignment  of either 
${\frac{1}{2}^+}$ or ${\frac{5}{2}^+}$ to the ground state of $^{17}$C
could enhance the partial cross sections to the ground state of $^{16}$C,
as it brings in the $\ell = 0$ component to this transition. However,
presence of this component has been ruled out in \cite{val00} from the
measurements of the corresponding parallel momentum distribution, which
is found to be broad and similar to the distribution of a $\ell =2$ state.
This excluded the $J^\pi = \frac{1}{2}^+$ assignment to the $^{17}$C 
ground state. At the same time,
with $J^{\pi} = \frac{5}{2}^+$, the calculated partial cross sections
were found to be in
disagreement with the data \cite{val00}. It has been concluded 
\cite{val00} that only with the assignment $J^\pi = \frac{3}{2}^+$
for the ground state of $^{17}$C, the trend of the experimental partial
cross sections for the excited states can be explained.

In our case the situation is similar. Of course, here 
the data for the parallel momentum distribution corresponding
to the ground state of $^{16}$C, have larger statistical errors
(see Fig. 2), which may not allow an unambiguous assignment of the
$\ell$-value to this state. However, with the $J^\pi$ value of 
$\frac{1}{2}^+$ for the ground state of $^{17}$C and the corresponding
spectroscopic factors \cite{war92}, the pure Coulomb partial
cross sections are found to be 690 mb, 40 mb and 89 mb corresponding
to the states with excitation energies of 0.0 MeV, 1.7 MeV and 4.1 MeV,
respectively. Looking at the data, these cross sections, on their own,   
rule out this assignment of the spin-parity to the $^{17}$C ground state.
Consideration of the nuclear partial cross sections will worsen the
comparison with the data even further. On the other hand,
with $J^\pi = {\frac{5}{2}}^+$,
the theoretical partial cross section for the transition to
the ground state ($\sim$ 170 mb) is closer to the data. However,
the cross section to the group of states at 4.1 MeV is predicted to be 
larger than that to the 1.77 MeV 2$^+$ state, which is in disagreement
with the pattern of the experimental data. Therefore, our calculations too 
suggest that the most suitable spin-parity assignment for the ground state
of $^{17}$C is $\frac{3}{2}^+$. 
   
Thus, the discrepancy between theoretical and experimental
cross sections for the transition to the ground state of $^{16}$C
can not be resolved by a different spin-parity assignment to the
$^{17}$C ground state. Furthermore, 
the theories of the Coulomb and nuclear breakup reactions used  
here are quite robust; they have been used earlier to explain
successfully the Coulomb and nuclear dominated data as 
discussed above. Although it is desirable to calculate both these
cross sections on an equal footing within the same quantum mechanical
theory, which would also include the Coulomb-nuclear interference terms,
yet this is unlikely to explain the observed one order of magnitude
difference between the data and the calculations for the $^{16}$C
ground state transition. To solve this problem, a couple of possibilities
are discussed in Ref. \cite{val00}. They essentially, try to invoke  
mechanisms which go beyond the simple direct one-neutron removal
process assumed in the present analysis. If the $\frac{5}{2}^+$
state lies very close to the $\frac{3}{2}^+$ ground state of $^{17}$C,
then the coupled channel calculations for the breakup process may help in
resolving this discrepancy. However, such calculations, which are 
difficult to perform at these high energies, have not yet been carried out.  
 
The parallel momentum distributions (PMD) of each of the $^{16}$C
core states, are shown in Fig. 2. In this figure we have shown only the
pure Coulomb calculations. As stated earlier,
the data have large statistical errors for the ground state transition.
Even then, it appears to be a broad distribution which supports
a $d$-wave relative motion for the core(g.s.)-neutron relative motion.
For the excited states, the momentum
distribution data can only be
understood by a combination of the $s$-wave and $d$-wave distributions,
which have narrow and broad widths respectively.
However, the difference in relative contributions of these components 
for the excited states at 1.77 MeV and 4.1 MeV should be noted. This
can be attributed to the relative difference in the  
spectroscopic factors for $\ell$ = 0 and 2 configurations in two cases.
The fact that with values of $C^2S$ (as given table I)
the shapes of the experimental PMDs are described rather well by the theory,
provides further support to the assignment of a spin-parity of
${\frac{3}{2}}^+$ to the $^{17}$C ground state, as these 
spectroscopic factors are based upon this assumption.
This conclusion is unlikely to be affected by the inclusion of the nuclear
breakup in the calculations, as the shapes of the corresponding PMDs
are not different from those of the Coulomb ones shown in this
figure.
\begin{figure}
\begin{center}
\mbox{\epsfig{file=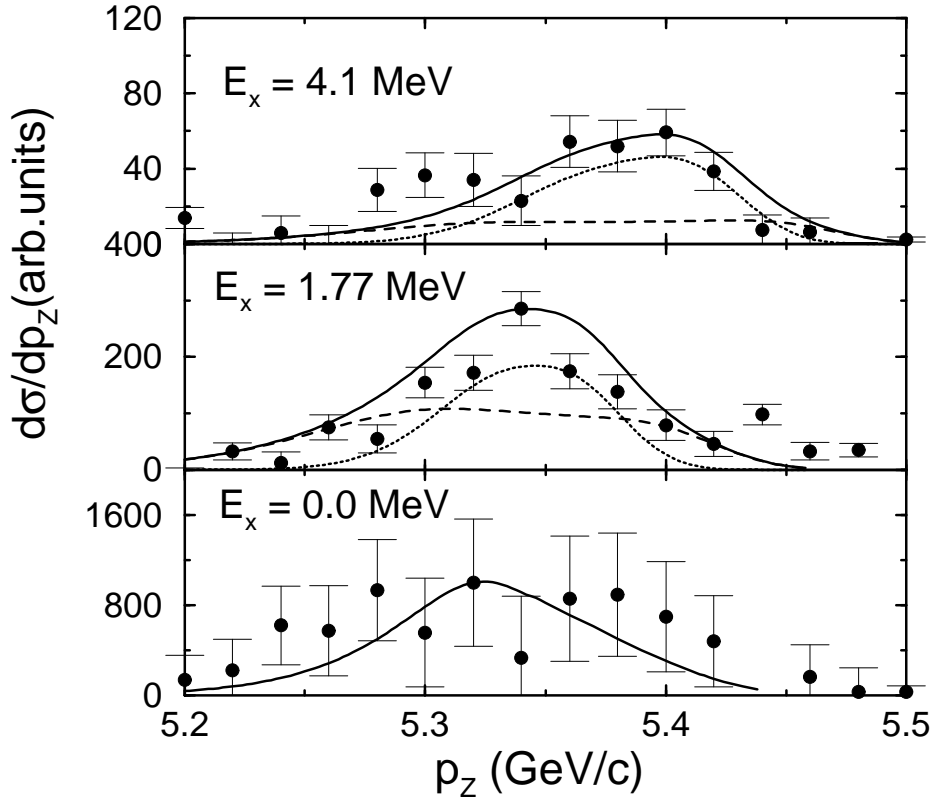,height=10.0cm}}
\end{center}
\vskip .3in
\caption {
Comparison of the pure Coulomb breakup calculations and the 
experimental data for the partial parallel momentum distributions
of the $^{16}$C core states with the excitation energy (E$_x$) as
indicated. The dotted and dashed lines represent the
results of the calculations performed with 
$s$-wave and $d$-wave configurations for the 
core - valence neutron system respectively, while the solid line
represents their sum.
} 
\label{fig:figb}
\end{figure}
\noindent
In summary, we investigated the role of the Coulomb and nuclear breakup
mechanisms in the one-neutron removal reaction
$^{197}$Au($^{17}$C,$^{16}$C$\gamma$)X studied at the beam energy of
61 MeV/nucleon. Partial cross sections and parallel momentum
distributions were measured for the ground as well as excited bound
states of the core fragment $^{16}$C, by detecting the core residues
in coincidence with the $\gamma$ rays emitted in the decay of the
excited core states. Substantial partial cross sections were found 
for the excited core states, which is in contrast to the results seen
in the case of the similar measurement performed with the one-neutron
halo nucleus $^{14}$B on this target.

An important observation of our study is that while
for the excitation of the core ground state 
the nuclear partial cross sections are of the similar magnitude as the 
Coulomb ones, they dominate the calculated cross sections 
for transitions to the excited states.
The sum of these two cross sections, weighted by
the spectroscopic factors taken from Ref. \cite{war92} (which is 
based on a ${\frac{3}{2}}^+$ spin-parity assignment for the ground 
state of $^{17}$C), is able to provide a good description of the
experimental data for the partial cross sections for transitions to
the excited states of the core. The dominance of the  
nuclear breakup effects in these data (taken on  
a heavy target), is reminiscent of the similar observations
made \cite{baur84} in the case of the breakup of stable nuclei.
This supports the fact that $^{17}$C is not a halo nucleus even
though it has a small one-neutron separation energy.
However, the theory is unable to describe 
the data for the transition to the core ground state.  
Similar observations were made 
earlier in the measurement of this reaction on a $^9$Be target.
This situation can not be remedied by a different spin-parity
assignment for the $^{17}$C ground state.

The shapes of the parallel momentum distributions are  
described well by the theory. For the excited states an admixture 
of the $s$-wave and $d$-wave configurations with spectroscopic factors
as given above is necessary to explain the shapes of the observed
PMDs. Our work underlines the need for a proper quantum mechanical 
calculation of the nuclear and Coulomb-nuclear interference breakup
terms. 

Authors are thankful to other members of the experimental
group, namely, T. Aumann, D. Bazin, J.A. Caggiano,
B. Davids, T. Glasmacher, P.G. Hansen, R.W. Ibbotson, A. Navin, B.V.
Pritychenko, H. Scheit, B.M. Sherrill, M. Steiner, and 
J. Yurkon, for their valuable collaboration during the experiment and for
letting them use the data presented in this paper.
Thanks are also due to Alex Brown for providing the spectroscopic
factors and to Jeff Tostevin for the eikonal model
code used here to perform the nuclear breakup calculations. One of the
authors (RS) would like to express his sincere thanks to Pawel Danielewicz
for his kind hospitality in the theory group of the Cyclotron Laboratory
of the Michigan State University and to Gregers 
Hansen for several very useful and illuminating discussions. 
This work has been supported by the National Science Foundation under
Grant PHY-0070818.

\end{document}